\documentclass[reprint,amsmath,amssymb,aps,prb,superscriptaddress,nofootinbib]{revtex4-2}

\usepackage{graphicx}
\usepackage{dcolumn}
\usepackage{bm}
\usepackage{color}
\usepackage{ulem}
\usepackage{comment}
\newcommand{\ua}{\uparrow}
\newcommand{\da}{\downarrow}
\newcommand{\dg}{\dagger}

\usepackage{CJKutf8}

\begin{document}


\title{Possibility of BCS-BEC crossover in $\kappa$-type organic superconductors}


\author{Hiroshi Watanabe}
\email{watanabe.hiroshi@nihon-u.ac.jp}
\affiliation{
Department of Liberal Arts and Basic Sciences, College of Industrial Technology, Nihon University, Chiba 275-8576, Japan
}
\affiliation{
Research Organization of Science and Technology, Ritsumeikan University, Shiga 525-8577, Japan
}
\author{Hiroaki Ikeda}
\affiliation{
Department of Physical Science, Ritsumeikan University, Shiga 525-8577, Japan
}


\date{\today}

\begin{abstract}
The realization of BCS-BEC crossover in superconductors, which smoothly connects Bardeen-Cooper-Schrieffer (BCS) theory with Bose-Einstein Condensation (BEC) in fermion systems, is an intriguing recent topic in strongly correlated electron systems.
The organic superconductor $\kappa$-(BEDT-TTF)$_4$Hg$_{2.89}$Br$_8$ ($\kappa$-HgBr) under pressure is one of the leading candidates, owing to its unique metallic spin-liquid nature and tunable electron correlation.
We theoretically investigate the extended Hubbard model for $\kappa$-HgBr and discuss the possibility of the BCS-BEC crossover by systematically calculating superconducting correlation function, coherence length, superfluid weight, and chemical potential.
Our findings show that the BCS-BEC crossover can be observed when competing phases, such as Mott insulators and charge and/or spin orders, are suppressed by appropriate hole doping.
$\kappa$-HgBr is just the case because both the Mott insulating phase and magnetic orders are absent due to its nonstoichiometric Hg composition and geometrical frustration.
We further propose that other $\kappa$-type organic superconductors could serve as potential candidates of the BCS-BEC crossover if their band fillings and degree of geometrical frustration are systematically tuned.

\end{abstract}


\maketitle


\section{Introduction}
Superconductivity (SC), even over a century after its discovery, continues to present compelling research opportunities in condensed matter physics. 
Possibility of a crossover from BCS (Bardeen-Cooper-Schrieffer) type to BEC (Bose-Einstein condensate) type behavior~\cite{Eagles,Leggett,Nozieres,Randeria1,Chen1,Strinati} in solid-state materials is one of the recent intriguing topics.
Although the BCS-BEC crossover has been established in ultracold atom systems~\cite{Greiner,Regal,Randeria2}, most SC in solid-state materials is considered to reside in the BCS regime~\cite{Yanase,Sous};
criteria for BCS-BEC crossover, such as $\Delta / E_{\text{F}} \sim 1$ (superconducting gap $\Delta$ comparable to the Fermi energy $E_{\text{F}}$) or $k_{\text{F}}\xi \sim 1$ (coherence length $\xi$ comparable to the average interparticle distance $\sim 1/k_{\text{F}}$), are typically not met.
Recently, however, various strongly correlated materials, including iron-based superconductors~\cite{Kasahara,Rinott,Boker,Hashimoto}, organic superconductors~\cite{Suzuki,Imajo}, two-dimensional gated semiconductors~\cite{Nakagawa}, and even cuprate superconductors~\cite{Chen2}, have been proposed to lie in the BCS-BEC crossover region.
In addition to meeting the criteria mentioned above, these materials exhibit pseudogap behavior~\cite{Nakagawa,Imajo} or flattened band dispersions~\cite{Hashimoto}, both of which are characteristic features of the BCS-BEC crossover. 

In strongly correlated electron systems, realizing BCS-BEC crossover poses several challenges:
(i) The range over which interactions can be controlled by physical or chemical pressure is limited compared with the Feshbach resonance used for ultracold atoms.
(ii) Competing phases, such as Mott insulators and charge and/or spin orders, are often present.
These competing phases generally mask SC, preventing its realization.
(iii) Multiorbital effects, which frequently produce nontrivial phenomena in strongly correlated electron systems, add further complexity, making the identification of the BCS-BEC crossover challenging~\cite{Rinott,Boker,Hashimoto,Reyes,Yerin,Tajima,Witt}.

Despite these challenges, the quasi-two-dimensional organic superconductor $\kappa$-(BEDT-TTF)$_4$Hg$_{2.89}$Br$_8$ ($\kappa$-HgBr)~\cite{Lyubovskaya1,Lyubovskaya2} under pressure has emerged as a promising candidate for realizing BCS-BEC crossover.
Owing to the tunability of its lattice constant and bandwidth by pressure, electron correlation in $\kappa$-HgBr can be controlled more easily than in inorganic materials.  
The superconducting transition temperature $T_{\text{c}}$ shows a dome-shaped pressure dependence and non-Fermi-liquid behavior is observed on the low-pressure (strongly correlated) side~\cite{Oike1,Oike2}, both of which indicate the potential for BCS-BEC crossover near the $T_{\text{c}}$ peak.
The estimated in-plane coherence length $k_{\text{F}}\xi_{\parallel}$ ranges from approximately 3 (BEC-like at 0.24 GPa) to 50 (BCS-like at 1.0 GPa)~\cite{Suzuki}, further suggesting a pressure-induced BCS-BEC crossover.
$\kappa$-HgBr is also notable as a candidate for a doped quantum spin liquid (QSL)~\cite{Oike2}, exhibiting metallic properties due to its nonstoichiometric Hg composition of 2.89. 
The absence of neighboring magnetic orders and Mott insulators is favorable for realizing BCS-BEC crossover.
Additionally, the band structure around the Fermi energy is relatively simple and free from complex multiorbital effects, enhancing its suitability for this crossover.

As discussed above, $\kappa$-HgBr is considered an ideal system for realizing BCS-BEC crossover.
In this paper, we theoretically investigate the possibility of BCS-BEC crossover in $\kappa$-HgBr and related organic superconductors.
We introduce a four-band extended Hubbard model as an effective low-energy model and analyze the ground state property of this model by varying electron correlation and band filling using the variational Monte Carlo (VMC) method.
We calculate the superconducting correlation function, coherence length, superfluid weight, and chemical potential as criteria for BCS-BEC crossover.
Our results indicate that the BCS-BEC crossover can be observed at a hole doping rate of $\delta=0.06$, where SC persists even in the strongly correlated region due to the absence of competing orders.
We also demonstrate transitions to a Mott insulator ($\delta=0$) and a charge-ordered (stripe) phase ($\delta=0.11$), both of which mask the BCS-BEC crossover.
Our findings align with the observed behavior of real materials and support the feasibility of the BCS-BEC crossover in $\kappa$-HgBr and related organic superconductors.

\begin{figure}[b]
\centering
\includegraphics[width=1.0\hsize]{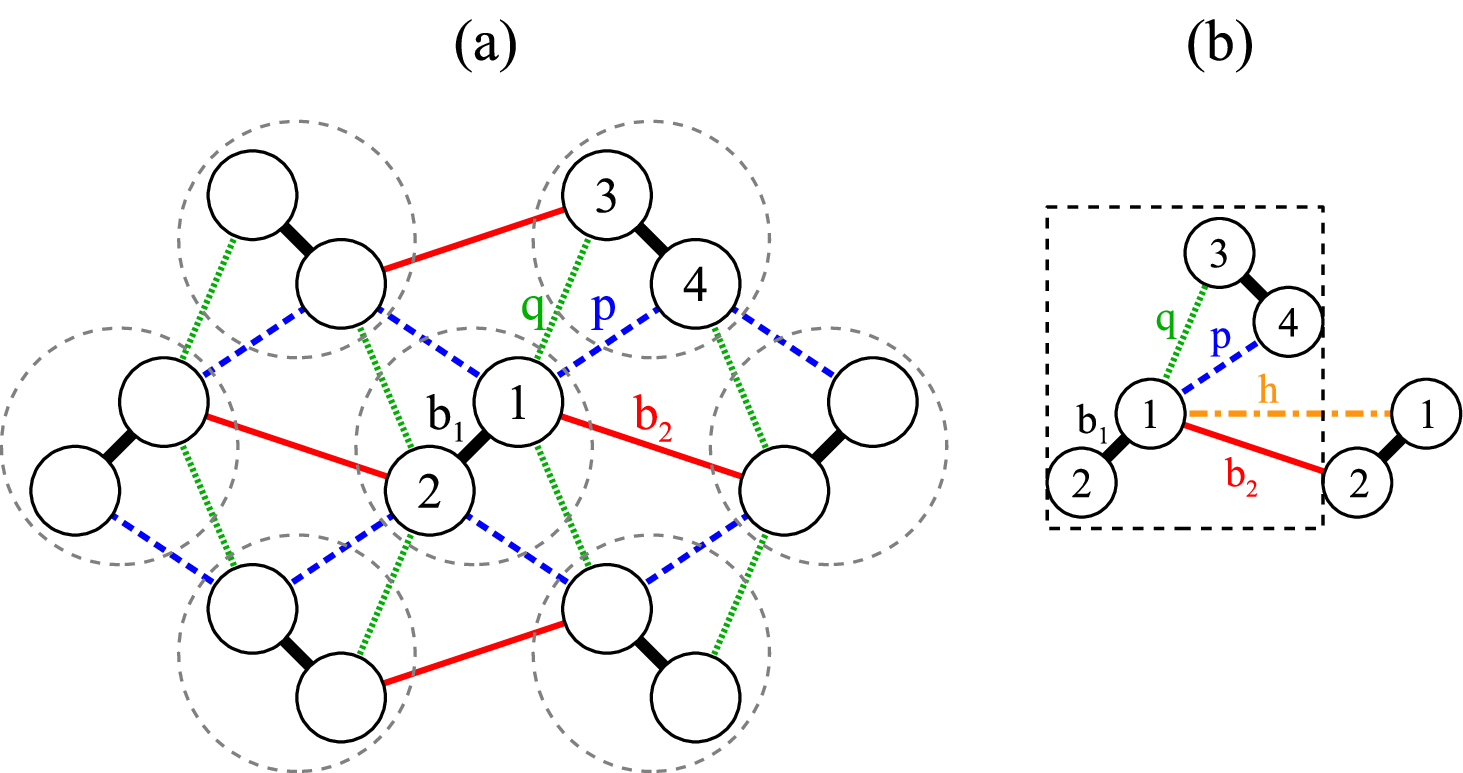}
\caption{\label{lattice}
(a) Schematic lattice structure of the $\kappa$-type system.
Solid circles represent BEDT-TTF molecules connected by $b_1$, $b_2$, $p$, and $q$ bonds.
Dashed circles indicate dimers, whose centers form an anisotropic triangular lattice. 
(b) The unit cell (dashed rectangle) contains four molecules labeled 1, 2, 3, and 4.
The $h$ bond exhibits the largest gap for an extended-$s$+$d_{x^2-y^2}$ type gap function.
}
\end{figure}

\section{Model and Method}
Figure~\ref{lattice}(a) shows the schematic lattice structure of the $\kappa$-type system.
Molecules connected by the $b_1$ bond form dimers (dashed circles) and the centers of these dimers form an anisotropic triangular lattice structure.
These dimers are often treated as single units in simplified models, resulting in the one-band Hubbard model on an anisotropic triangular lattice~\cite{Kino}.
This ``dimer model'' has been extensively studied and has successfully described many aspects of the Mott transition in these materials.
However, recent studies have revealed the significance of intradimer molecular degrees of freedom, which can give rise to nontrivial and intriguing effects on superconducting, charge, and spin properties~\cite{Seo,Kuroki,Hotta,Naka1,Naka2}.
To address these issues, we adopt the original four-band description, where the unit cell contains four molecules.

Here, we study the four-band extended Hubbard model defined as~\cite{Watanabe1,Watanabe2}
\begin{equation}
H=-\sum_{\left<i,j\right>\sigma}t_{ij}(c^{\dg}_{i\sigma}c_{j\sigma}+\text{H.c.})
+U\sum_{i}n_{i\ua}n_{i\da}
+\sum_{\left<i,j\right>} V_{ij}n_in_j, \label{Hamiltonian}
\end{equation}
where $c^{\dg}_{i\sigma}$ ($c_{i\sigma}$) is a creation (annihilation) operator for an electron at the $i$-th molecular site with spin $\sigma$.
$n_{i\sigma}=c^{\dg}_{i\sigma}c_{i\sigma}$ and $n_i=n_{i\ua}+n_{i\da}$ are the number operators, representing the electron number for each spin and total electron number, respectively.
$U$ and $V_{ij}$ denote the on-site and intersite Coulomb interactions, respectively.
$\left<i,j\right>$ denotes a pair of neighboring molecules in the $\kappa$-type geometry, labeled by $b_1$, $b_2$, $p$, and $q$.

A set of transfer integrals $t_{ij}$ is estimated from a first-principles band calculation.
However, performing a band calculation for $\kappa$-HgBr is challenging due to its nonstoichiometric Hg composition.
Instead, we use band calculation results for $\kappa$-(BEDT-TTF)$_2$Cu$_2$(CN)$_3$ ($\kappa$-CN)~\cite{Koretsune}, which is also a candidate for a QSL~\cite{Shimizu}.
Both $\kappa$-CN and $\kappa$-HgBr share nearly isotropic triangular lattice structures, leading to strong geometrical frustration and the suppression of magnetic long-range order despite strong electron correlation.
Moreover, the temperature dependence of the magnetic susceptibility in $\kappa$-HgBr shows behavior similar to $\kappa$-CN~\cite{Oike2}, further supporting their shared spin dynamics. 
The obtained parameters are ($t_{b_1},t_{b_2},t_p,t_q)=(199, 91, 85, -17$) meV = (1, 0.457, 0.427, -0.085) $t_{b_1}$.
In the following, the largest transfer integral $t=t_{b_1}$ is set as the unit of energy.
The unit cell contains four molecules [dashed rectangle in Fig.~\ref{lattice}(b)], resulting in a four-band electronic structure.
The noninteracting band structure is shown in Fig.~\ref{band-FS}(a) in Appendix~\ref{A-A}.
The initial electron density per molecular site is 3/4-filling, or 1/4-filling in a hole picture, which corresponds to 1/2-filling when each dimer is considered a unit (one hole per dimer).
To investigate the possibility of BCS-BEC crossover in the $\kappa$-type system, we vary the band filling by hole doping and calculate physical quantities relevant to SC.

We employ a VMC method~\cite{McMillan,Ceperley,Yokoyama} to treat the effects of Coulomb interactions.
The trial wave function used here is of the Gutzwiller-Jastrow type, $\left|\Psi\right>=P_{J_\text{c}}P_{J_\text{s}}\left|\Phi\right>$, where $\left|\Phi\right>$ is a one-body state constructed by diagonalizing the one-body Hamiltonian, which includes the superconducting gap $\tilde{\Delta}_{ij}$, renormalized transfer integrals $\tilde{t}_{ij}$, and renormalized chemical potential $\tilde{\mu}$.
$P_{J_\text{c}}$ and $P_{J_\text{s}}$ represent the charge and spin Jastrow factors, incorporating $v^{\text{c}}_{ij}$ and $v^{\text{s}}_{ij}$, respectively.
The explicit form of $\left|\Psi\right>$ are shown in Appendix~\ref{A-B}.
The variational parameters in $\left|\Psi\right>$ are optimized simultaneously using the stochastic reconfiguration method~\cite{Sorella,Yunoki}.
Once $\left|\Psi\right>$ is determined, various physical quantities can be calculated.

\begin{figure*}
\includegraphics[width=1.0\hsize]{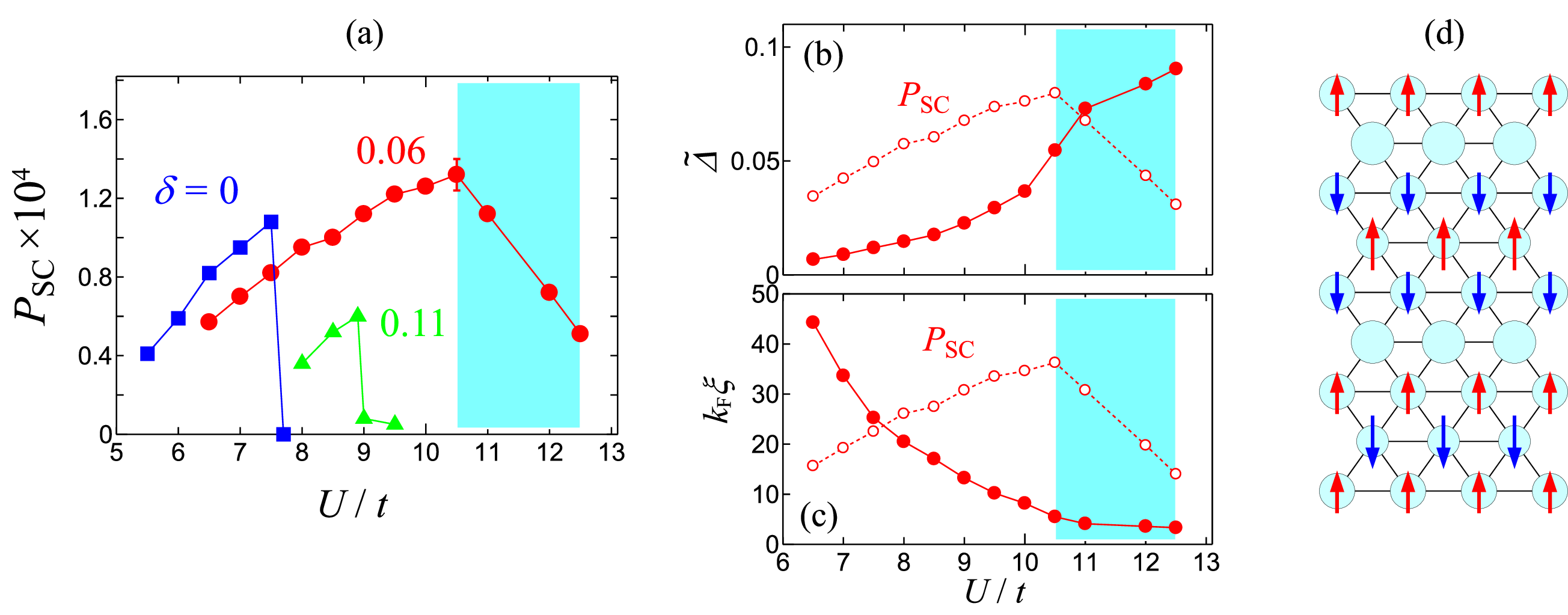}
\caption{\label{PSC}
(a) $U/t$ dependence of the superconducting correlation function $P_{\text{SC}}$ for hole doping rates $\delta$=0, 0.06, and 0.11.
The corresponding hole densities are 576/576, 608/576, and 640/576 (holes/dimers), respectively, with $N_{\mathrm{S}}=24\times24=576$.
The blue-shaded area indicates the BCS-BEC crossover region.
The statistical errors from Monte Carlo sampling are within the symbol size, except for $U/t=10.5$.
Around $U/t=10.5$, fluctuations between BCS- and BEC-type states lead to relatively large errors, indicated by the error bar. 
(b) $U/t$ dependence of the superconducting gap $\tilde{\Delta}$.
(c) $k_{\text{F}}\xi$ as a function of $U/t$.
(d) Charge and spin configuration of the C4S8 stripe phase stabilized for $U/t>8.9$ at $\delta=0.11$.
Circles (arrows) represent the total hole (spin) density in each dimer, with radius (length) proportional to the hole (spin) density.
}
\end{figure*}

\section{Result}
\subsection{Superconducting correlation function}
First, we examine the pair correlation function defined as
\begin{equation}
P_{\alpha}(\textbf{r})=\frac{1}{N_{\mathrm{S}}}
\sum_i\bigl<\Delta^{\dagger}_{\alpha}(\textbf{R}_i) \Delta_{\alpha}(\textbf{R}_i+\textbf{r})\bigr>,
\end{equation}
where $i$ runs over all molecular sites, $N_{\mathrm{S}}$ is the number of dimers, and $\Delta^{\dagger}_{\alpha}(\textbf{R}_i)$ represents the creation operator for singlet pairs on molecules connected by the $\alpha$ bond,
\begin{equation}
\Delta^{\dagger}_{\alpha}(\textbf{R}_i)
=(c^{\dagger}_{i\uparrow}c^{\dagger}_{i+\alpha\downarrow}+c^{\dagger}_{i+\alpha\uparrow}c^{\dagger}_{i\downarrow})/\sqrt{2}.
\end{equation}
If $P_{\alpha}(\textbf{r})$ converges to a finite value as $|\textbf{r}|=r\rightarrow\infty$, superconducting long-range order is present.
Indeed, $P_{\alpha}(\textbf{r})$ shows good convergence for $r\gtrsim 4$, and we take an average up to $r \approx 10$ to define it as $P_{\alpha}$.
We calculate $P_{\alpha}$ up to the 15th-neighbor bond ($b_1$, $b_2$, $p$, $q$, $h$, $\dots$), with the largest value taken as the superconducting correlation function $P_{\text{SC}}=P_{\alpha_{\text{max}}}$.
Our results confirm that the symmetry of the superconducting gap function is of the extended-$s$+$d_{x^2-y^2}$ type~\cite{Powell,Kuroki,Guterding,Watanabe1,Zantout,Watanabe2}, and the $h$ bond exhibits the largest gap in real space.
Further details on the gap function are provided in Appendix~\ref{A-C}.

Figure~\ref{PSC}(a) shows the $U/t$ dependence of $P_{\text{SC}}$ for different hole doping rates $\delta$.
The ratio between $U$ and intersite Coulomb interactions is fixed as ($V_{b_1}$,$V_{b_2}$,$V_p$,$V_q$)/$U$=(0.5, 0.28, 0.33, 0.29), assuming a $1/r$-dependence~\cite{Koretsune,Nakamura,Yoshimi}.
For $\delta=0$ (576 holes/576 dimers), $P_{\text{SC}}$ increases with increasing $U/t$, displaying BCS-like behavior, and then drops discontinuously to approximately zero at $U/t\approx7.7$ due to the Mott transition (from SC to a Mott insulator).
This corresponds to the pressure-induced Mott transition generally observed in the $\kappa$-type system~\cite{Kanoda}. 
For $\delta=0.11$ (640 holes/576 dimers), $P_{\text{SC}}$ shows similar behavior to the $\delta=0$ case.
At $U/t\approx 8.9$, a transition to a charge-ordered phase occurs and $P_{\text{SC}}$ decreases discontinuously.
The ordering pattern is analogous to the stripe phase observed in cuprate superconductors at $\delta\approx1/8=0.125$~\cite{Tranquada,Watanabe3}.
In this case, the charge periodicity is four and the spin periodicity is eight (C4S8) when each dimer is considered a unit, as shown in Fig.~\ref{PSC}(d).
Given the similarity to cuprate superconductors~\cite{McKenzie}, it is natural that the stripe phase also appears in the $\kappa$-type system at $\delta\approx1/8$.

In contrast, significantly different behavior is observed for $\delta=0.06$ (608 holes/576 dimers).
Here, $P_{\text{SC}}$ increases with increasing $U/t$ and then gradually decreases for $U/t > 10.5$ without any phase transitions.
The optimized superconducting gap $\tilde{\Delta}=\tilde{\Delta}_h$ for $\delta=0.06$ is also shown in Fig.~\ref{PSC}(b).
While $P_{\text{SC}}$ exhibits a peak structure, $\tilde{\Delta}$ increases monotonically with $U/t$.
For $U/t>10.5$, strong electron correlation suppresses the coherence of Cooper pairs, leading to a decrease in $P_{\text{SC}}$, which is a ``true'' order parameter of SC ($\sim T_{\text{c}}$), as $U/t$ increases.
Meanwhile, $\tilde{\Delta}$ reflects a local superconducting correlation and is continuously enhanced by electron correlation.
This observed inverse relationship---where local pairing gaps increase as long-range pair coherence decreases---serves as a hallmark of the BCS-BEC crossover.
We also plot $k_{\text{F}}\xi=k_{\text{F}}\cdot \hbar v_{\text{F}}/(\pi \tilde{\Delta})$~\cite{Paramekanti} in Fig.~\ref{PSC}(c).
$k_{\text{F}}\xi$ decreases monotonically with $U/t$ and saturates to approximately $O(1)$, where the coherence length $\xi$ is comparable to the average interparticle distance $\sim1/k_{\text{F}}$.
This behavior reproduces the estimated $k_{\text{F}}\xi$ for $\kappa$-HgBr~\cite{Suzuki} and supports the occurrence of BCS-BEC crossover.

Since the conventional understanding of the BCS-BEC crossover is associated with attractive interactions, it is essential to clarify the mechanism of SC and the BCS-BEC crossover in the presence of the repulsive interactions, specifically the strong local Coulomb interactions parametrized by $U/t$ in this study.
As $U/t$ increases, the spins of neighboring electrons tend to align antiparallel to gain the exchange energy, thereby enhancing local superconducting correlation alongside antiferromagnetic correlation, even in the presence of globally repulsive interactions.
In other words, local repulsive interactions effectively generate attractive interactions between neighboring electrons.
This mechanism induces various competing phases, including SC, charge and/or spin order, and other nontrivial quantum phases.
It is a well-established concept in strongly correlated systems, particularly in high-$T_{\mathrm{c}}$ cuprates.

The reason why the BCS-BEC crossover is observable at $\delta=0.06$ is that the competing phases are suppressed at this doping level.
In square lattice models, SC is often masked by other ordered phases, such as charge and/or spin order, when $U/t$ is large.
However, due to the geometrical frustration inherent to a triangular lattice structure, such competing orders are suppressed in the present model.
Consequently, SC persists in a strongly correlated region where $U/t$ is sufficiently large to realize the BCS-BEC crossover.
For $U/t>12.5$, an inhomogeneous state emerges, where the variational energy depends on the initial configuration of the Monte Carlo sampling.
In this region, phase separation between SC and competing orders, such as Mott insulators or short- and long-range charge-ordered phases, is expected.
Indeed, inhomogeneous SC has been observed in $\kappa$-HgBr at low-pressure (strongly correlated conditions)~\cite{Oike1}.
This inhomogeneity results from strong electron correlation effects and resembles the behavior seen in the low-doping regions of cuprate superconductors~\cite{Emery}.

\subsection{Superfluid weight}
\begin{figure}
\centering
\includegraphics[width=1.0\hsize]{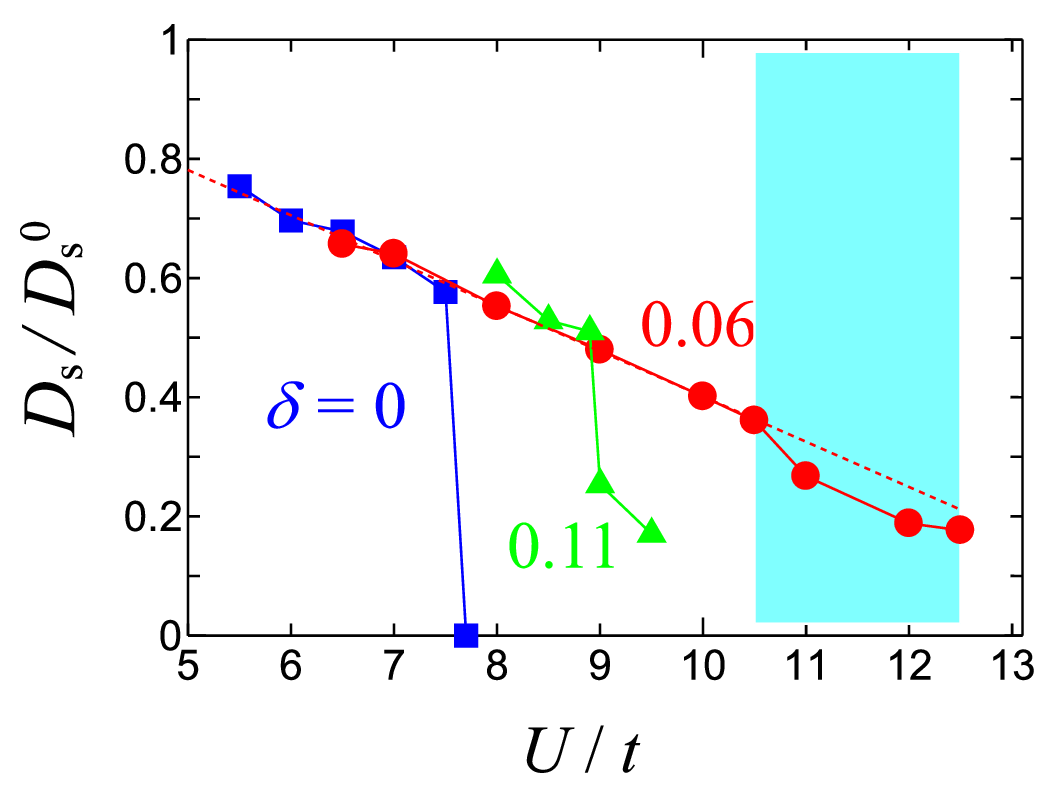}
\caption{\label{Ds}
$U/t$ dependence of the normalized superfluid weight $D_{\text{s}}/D_{\text{s}}^0$ for $\delta=0$, 0.06, and 0.11.
$D_{\text{s}}^0$ is the value at $U/t=0$ for each $\delta$.
The red-dotted line represents the fit for $U/t< 10.5$ and its extrapolation to $U/t> 10.5$. 
The blue-shaded area indicates the BCS-BEC crossover region.
}
\end{figure}

We also examine the superfluid weight $D_{\text{s}}\sim n_{\text{s}}/m^*$, where $n_{\text{s}}$ is the superfluid density and $m^*$ is the effective electron mass. 
$D_{\text{s}}$ is calculated by adding the vector potential $\textbf{A}$ to the Hamiltonian [Eq.~\eqref{Hamiltonian}]~\cite{Millis,Scalapino,Hetenyi,Tamura} (see Appendix~\ref{A-D} for details of the calculation).
Figure~\ref{Ds} shows the $U/t$ dependence of $D_{\text{s}}/D_{\text{s}}^0$ for $\delta=0$, 0.06, and 0.11, where $D_{\text{s}}^0$ is the value at $U/t=0$ for each $\delta$.
For $\delta=0$, $D_{\text{s}}/D_{\text{s}}^0$ decreases with $U/t$ and goes to zero at the Mott transition.
For $\delta=0.11$, $D_{\text{s}}/D_{\text{s}}^0$ also decreases with $U/t$ but jumps to a small, finite value at the first-order transition to the stripe phase, as the electron mobility is suppressed but remains finite.
For $\delta=0.06$, however, $D_{\text{s}}/D_{\text{s}}^0$ decreases with $U/t$ and shows no discontinuous change across $U/t\approx 10.5$, where $P_{\text{SC}}$ exhibits a peak structure, as shown in Fig~\ref{PSC}(a).
In the BCS regime, all particles contribute to SC, and $n_{\text{s}}$ remains constant.
Consequently, $D_{\text{s}}/D_{\text{s}}^0$ is proportional to $m/m^* \sim z$, where $z$ is the quasiparticle renormalization factor.
Indeed, the dependence observed in Fig.~\ref{Ds} aligns with that of $z$ for moderate $U/t$ in previous studies~\cite{Merino,TWatanabe}, suggesting that the decrease in $D_{\text{s}}/D_{\text{s}}^0$ is due to an increase in $m^*$. 
The deviation from the fit for $U/t>10.5$ suggests that the additional decrease results from a reduction in $n_{\text{s}}$, 
as some particles become incoherent and no longer contribute to SC.
This observation also supports the occurrence of the BCS-BEC crossover.
We believe that the reduction in $P_{\text{SC}}$ likewise results from the decrease in $n_{\text{s}}$.

The ratio $n_{\text{s}}/m^*$ can be estimated from the magnetic penetration depth $\lambda$, with $1/\lambda^2 \sim n_{\text{s}}/m^*$~\cite{Uemura}.  
The estimated $n_{\text{s}}/m^*$ for $\kappa$-HgBr at ambient pressure is $9.4\times 10^{54}$ (m$^3$kg)$^{-1}$, which is one order of magnitude smaller than those of undoped ($\delta=0$) $\kappa$-type superconductors, $\kappa$-(BEDT-TTF)$_2$Cu[N(CN)$_2$]Br ($\kappa$-Br) and $\kappa$-(BEDT-TTF)$_2$Cu(NCS)$_2$ ($\kappa$-NCS)~\cite{Wakamatsu}.
This low $n_{\text{s}}/m^*$ value for $\kappa$-HgBr indicates a heavy electron mass and reduced superfluid density.
We propose that $\kappa$-HgBr at ambient pressure lies in the BCS-BEC crossover region, where the effect of electron correlation is particularly strong. 

\begin{figure}
\centering
\includegraphics[width=1.0\hsize]{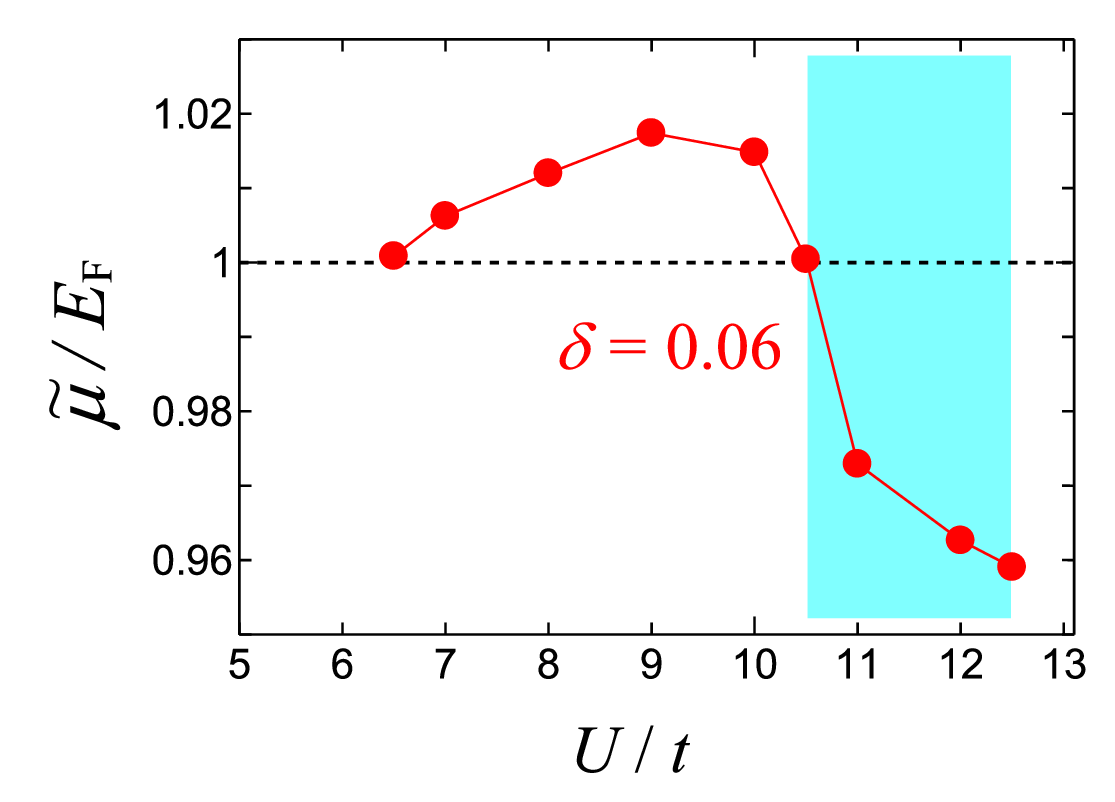}
\caption{\label{mu}
$U/t$ dependence of the renormalized chemical potential $\tilde{\mu}/E_{\text{F}}$ for $\delta=0.06$.
The blue-shaded area, where $\tilde{\mu}/E_{\text{F}}<1$, indicates the BCS-BEC crossover region.
}
\end{figure}

\subsection{Chemical potential}
The behavior of the fermionic chemical potential $\mu$ is one of the criteria for identifying the BCS-BEC crossover.
In this work, we use $\tilde{\mu}/E_\text{F}$ to evaluate the BCS-BEC crossover.
This approach relies on the variational parameter $\tilde{\mu}$ and the noninteracting Fermi energy $E_{\text{F}}$, which may deviate from the ``true'' values of $\mu$ and $E_{\text{F}}$.
Obtaining these true values would require additional calculations, which are beyond the scope of the present study.
We regard this as an important future problem to refine the quantitative understanding of the BCS-BEC crossover in strongly correlated systems.
Despite this limitation, we believe that the qualitative trends captured by $\tilde{\mu}/E_\text{F}$ provide meaningful insights.
Figure~\ref{mu} shows the $U/t$ dependence of $\tilde{\mu}/E_{\text{F}}$ for $\delta=0.06$.
Since the electronic carriers in this system are holes, both $\tilde{\mu}<0$ and $E_{\text{F}}<0$ are measured relative to the top of the band [see the band structure in Fig.~\ref{band-FS}(a) in Appendix~\ref{A-A}].
For a noninteracting case ($U/t=0$), $\tilde{\mu}/E_{\text{F}}=1$.
$\tilde{\mu}$ is renormalized due to the electron correlation and $\tilde{\mu}/E_{\text{F}}<1$ ($>1$) corresponds to a upward (downward) shift in Fig.~\ref{band-FS}(a).  

For $U/t<10.5$, $\tilde{\mu}/E_{\text{F}}$ gradually increases from unity as $U/t$ increases.
This behavior reflects electron correlation effects that enhance SC by expanding the Fermi surface area. 
However, for $U/t>10.5$, $\tilde{\mu}/E_{\text{F}}$ decreases and eventually falls below unity.
The decline in $\tilde{\mu}/E_{\text{F}}$ suggests that tightly bound pairs begin to decouple from the Fermi surface, leading to pseudogap formation~\cite{Chen1}.
It can be regarded as a precursor of BCS-BEC crossover.

In Fermi gas systems, the chemical potential $\mu$ typically approaches $\mu \lesssim 0$ in the BCS-BEC crossover regime.
However, in the present study, we find that $\mu$ remains on the order of $E_F$.
Meanwhile, the rapid suppression of the coherence length ($k_F\xi \sim 1$), the dome-shaped behavior of the superconducting correlation function $P_{\text{SC}}$, and the reduction of superfluid density suggest that the system enters the BCS-BEC crossover regime for $U/t>10.5$.

How, then, should we interpret the behavior of $\mu/E_F$?
The variational wave function obtained in this study exhibits striking similarities to that observed in cuprate superconductors near the Mott insulator~\cite{Watanabe4}, suggesting that strong electron correlations lead to electron localization.
In this regime, a significant fraction of electrons that originally contributed to the Fermi surface form incoherent localized spins, resulting in a substantial reduction in the number of effective mobile quasiparticles near the Fermi surface. 
Thus, while $\mu$ remains on the order of $E_F$ in this regime, the quasiparticles involved in superconductivity effectively behave as those in low carrier density systems.
In this context, the transition of $\mu/E_F$ from increasing to decreasing serves as evidence for the onset of the crossover, and the extent of its decrease corresponds to the effective number of mobile quasiparticles.
This behavior differs from that observed in the BCS-BEC crossover of Fermi gas systems and may instead represent a unique feature of the BCS-BEC crossover in repulsively interacting lattice systems.
A complete understanding of this phenomenon requires further detailed analysis, including precise estimations of the true values of $\mu$ and $E_F$.

\section{Conclusion}
Based on the behavior of $P_{\text{SC}}$, $k_{\text{F}}\xi$, $D_{\text{s}}/D_{\text{s}}^0$, and $\tilde{\mu}/E_{\text{F}}$, we conclude that the BCS-BEC crossover occurs at $\delta=0.06$ but not at $\delta=0$ or $\delta=0.11$ in the present model.
Appropriate hole doping suppresses competing orders, allowing the BCS-BEC crossover to emerge.
Most $\kappa$-type organic superconductors correspond to $\delta=0$, where the Mott transition takes place before reaching the BCS-BEC crossover region.
In contrast, $\kappa$-HgBr, with $\delta$ away from zero, has the potential to realize the BCS-BEC crossover.
Although the hole doping rate estimated from stoichiometry is $\delta=0.11$, we anticipate that the charge-ordered phase is suppressed by the effect of Hg$^{2+}$ ions in insulating layers (not included in the present model), which could eventually allow the BCS-BEC crossover to occur in $\kappa$-HgBr.

Our study reveals that the BCS-BEC crossover in repulsively interacting lattice systems exhibits distinct characteristics compared to that in Fermi gas systems.
While the chemical potential $\mu$ in Fermi gas systems typically approaches $\mu \lesssim 0$ in the crossover regime, we find that $\mu$ remains on the order of $E_\text{F}$ in our model.
We interpret this behavior as a consequence of strong electron correlation, which induces electron localization and significantly reduces the number of effective mobile quasiparticles.
As a result, even though $\mu$ does not decrease substantially, the system effectively behaves as a low carrier density system in the superconducting state.
This feature differs from conventional BCS-BEC crossover scenarios in Fermi gas systems and may represent a unique characteristic of the crossover in repulsively interacting lattice systems.

Establishing clear experimental criteria for identifying BCS-BEC crossover in strongly correlated electron systems remains an ongoing challenge and is one of the key motivations of this work.
The important parameters to control are $U/t$ and $\delta$.
Tuning $U/t$ can primarily be achieved by applying pressure, which increases the transfer integral $t$ and effectively reduces $U/t$.
If the BCS-BEC crossover occurs, one would expect to observe a dome-shaped transition temperature $T_{\text{c}}$, a rapid increase in coherence length $\xi$, and enhancements in superfluid weight $D_{\text{s}}$ and superfluid density $n_{\text{s}}$ with increasing pressure.
In addition, chemical pressure provides an alternative way to control $U/t$.
For instance, pseudogap behavior has been observed in the family of $\kappa$-(BEDT-TTF)$_2$X when the anion layer X is systematically substituted~\cite{Imajo}.
Utilizing both physical and chemical pressure enables access to a wide range of $U/t$, which is essential for identifying BCS-BEC crossover.

Additionally, hole doping may not directly alter $U/t$ but can suppress Mott insulating behavior, thereby enabling SC to persist in strongly correlated regimes.
While tuning $\delta$ in organic conductors is generally more challenging than in inorganic materials, recent advancements in the electric-double-layer transistor method~\cite{Kawasugi} have made it possible.
This method provides an alternative pathway, as undoped materials such as $\kappa$-CN may also achieve the BCS-BEC crossover through controlled doping.

Finally, we discuss the role of geometrical frustration.
As mentioned in the Introduction, suppressing competing phases is essential for realizing the BCS-BEC crossover.
In $\kappa$-CN and $\kappa$-HgBr, the nearly isotropic triangular lattice structure induces strong geometrical frustration, which suppresses magnetic ordering and allows SC to persist even in strongly correlated regimes.
This is not the case for $\kappa$-Br, $\kappa$-NCS, and $\kappa$-(BEDT-TTF)$_2$Cu[N(CN)$_2$]Cl ($\kappa$-Cl), where weaker geometrical frustration leads to antiferromagnetic ordering~\cite{Kanoda}.
However, it has been reported that the geometrical frustration can be controlled by combining uniaxial and hydrostatic pressures~\cite{Oike3}.
This approach could potentially enhance the geometrical frustration in $\kappa$-Br, $\kappa$-NCS, and $\kappa$-Cl, providing a pathway to achieve the BCS-BEC crossover in these materials.  

In summary, combining various experimental methods, including pressure tuning, chemical substitution, hole doping, and geometrical frustration control, offers a promising strategy to explore the BCS-BEC crossover in strongly correlated electron systems.
We hope this work will stimulate further experimental and theoretical investigations in this direction.

\begin{acknowledgments}
The authors thank K. Kanoda, H. Oike, H. Seo, and S. Imajo for useful discussions.
The computation has been done using the facilities of the Supercomputer Center, Institute for Solid State Physics, University of Tokyo.
This work was supported by a Grant-in-Aid for Scientific Research on Innovative Areas ``Quantum Liquid Crystals'' (KAKENHI Grant No. JP19H05825) from JSPS of Japan,
and also supported by JSPS KAKENHI (Grant Nos. JP20K03847, JP23H01130 and JP23K25827).
\end{acknowledgments}

\appendix
\section{Band Structure and Fermi surface}\label{A-A}
\begin{figure}
\centering
\includegraphics[width=1.0\hsize]{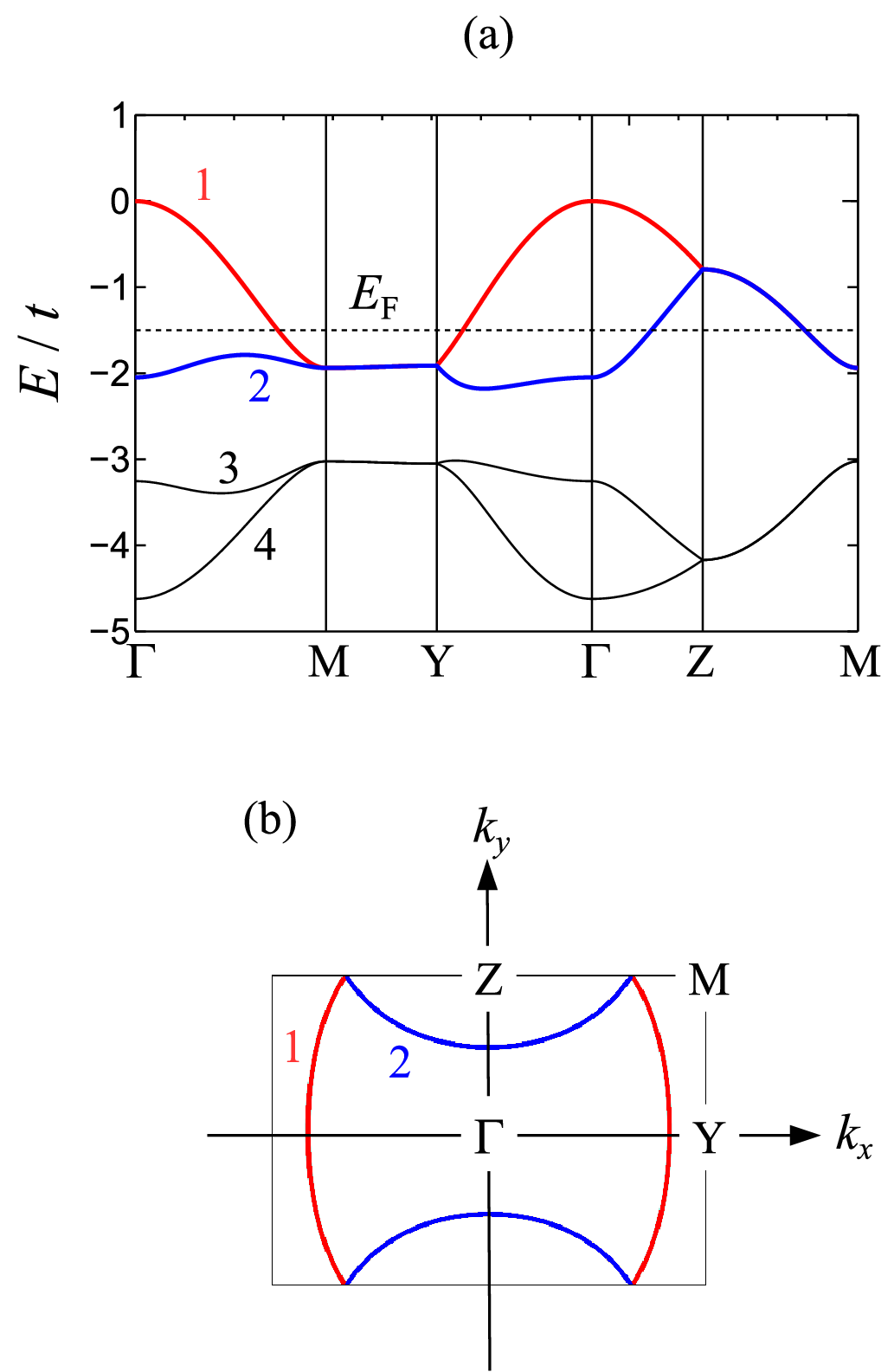}
\caption{\label{band-FS}
(a) Band structure and (b) Fermi surface for $\kappa$-HgBr.
High symmetry points are $\Gamma$(0,0), Y($\pi/R_x$), Z($\pi/2R_y$), and M($\pi/R_x,\pi/2R_y$).
}
\end{figure}

The noninteracting band structure and the Fermi surface in the present model are shown in Fig.~\ref{band-FS}.
The upper two bands (bands 1 and 2) contribute to the formation of the Fermi surface.
The unit cell in real space is $R_x\times 2R_y$ rectangle [Fig. 1(b)], where $R_y=0.7 R_x$.

\section{Trial Wave Function}\label{A-B}
The one-body part $\left|\Phi\right>$ of a trial wave function for superconductivity (SC) is obtained from the Bogoliubov de-Gennes (BdG) type Hamiltonian in real space,
\begin{equation}
    H_{\text{BdG}}=\sum_{i,j}\left(c^{\dg}_{i\ua}, c_{i\da}\right)
    \begin{pmatrix}
        T_{ij\ua} & \tilde{\Delta}_{ij} \\ 
        \tilde{\Delta}_{ji} & -T_{ji\da}
    \end{pmatrix}
    \begin{pmatrix}
        c_{j\ua} \\
        c^{\dg}_{j\da}
    \end{pmatrix},
\end{equation}
where $\sum_{i,j}$ runs over all molecular sites.
$T_{ij\sigma}$ represents the normal part, including renormalized transfer integrals $\tilde{t}_{ij}$ and renormalized chemical potential $\tilde{\mu}$.
Here, $\tilde{\mu}$ is a variational parameter and deviates from the Fermi energy $E_{\text{F}}$ due to electron correlation effects.
$\tilde{\Delta}_{ij}$ represents the superconducting gap in real space, forming the anomalous part of the Hamiltonian.
Therefore, the variational parameters to be optimized in $\left|\Phi\right>$ are $\tilde{t}_{ij}$, $\tilde{\mu}$, and $\tilde{\Delta}_{ij}$ with $\tilde{t}_{b_1}=t$ fixed as the unit of energy.
In this study, pairing interactions are considered between bonds up to the 15th neighbor (e.g., $b_1$, $b_2$, $p$, $q$, $h$, $\dots$).
Setting $\tilde{\Delta}_{ij}=0$ yields the paramagnetic phase.

The charge and spin Jastrow factors are defined as
\begin{equation}
P_{\text{J}_{\text{c}}}=\exp\Bigl[-\sum_{i,j}v^{\text{c}}_{ij}n_in_j\Bigr],
\end{equation}
and
\begin{equation}
P_{\text{J}_{\text{s}}}=\exp\Bigl[-\sum_{i,j}v^{\text{s}}_{ij}s^z_{i}s^z_{j}\Bigr],
\end{equation}
where $\sum_{i,j}$ runs over all molecular sites.
These factors control long-range charge and spin correlations, respectively.
The variational parameters to be optimized are $v^{\text{c}}_{ij}$ and $v^{\text{s}}_{ij}$.
Here, we do not include mean-field type charge and spin gaps, as they often lead to overestimations of charge and spin orders.
The Mott insulating and stripe phases observed in this study arise from the effects of $P_{\text{J}_{\text{c}}}$ and $P_{\text{J}_{\text{s}}}$.

\section{Superconducting Gap Function}\label{A-C}
In $\kappa$-type organic superconductors, the degree of geometrical frustration is an important parameter for both SC and magnetic orders.
In the dimer approximation, the ratio between the effective diagonal and horizontal interdimer transfer integrals is estimated as $|t'_{\text{eff}}/t_{\text{eff}}|\sim |t_{b_2}|/(|t_p|+|t_q|)$. 
A value of $|t'_{\text{eff}}/t_{\text{eff}}|\approx 1$ indicates strong geometrical frustration (triangular-lattice-like), whereas $|t'_{\text{eff}}/t_{\text{eff}}|\ll 1$ suggests weak geometrical frustration (square-lattice-like). 
The primary proposed symmetries for SC in $\kappa$-type organic superconductors are extended-$s$+$d_{x^2-y^2}$ and $d_{xy}$. 
The former symmetry is favored in systems with strong geometrical frustration, while the latter is favored in systems with weak geometrical frustration.
The competition between these symmetries is discussed in detail in our previous study~\cite{Watanabe2}.

The quantum spin liquid candidates $\kappa$-HgBr and $\kappa$-CN are expected to exhibit extended-$s$+$d_{x^2-y^2}$ symmetry due to their strong geometrical frustration.
In $\textbf{k}$-space representation, the main contribution of the gap function for the extended-$s$+$d_{x^2-y^2}$ type is given as~\cite{Watanabe2}
\begin{equation}
\begin{split}
    \Delta^{\alpha} &= \Delta^{\alpha}_1 \left[\cos\left(\frac{1}{2}k_xR_x + k_yR_y\right) + \cos\left(\frac{1}{2}k_xR_x - k_yR_y\right)\right] \\
    & + \Delta^{\alpha}_2 \cos k_xR_x,
\end{split}
\label{gap}
\end{equation}
where  $\alpha\,(=1,2)$ denotes the band index shown in Fig.~\ref{band-FS}.
The optimized $\tilde{\Delta}_{ij}$ in this study changes sign four times in real space, which is consistent with the extended-$s$+$d_{x^2-y^2}$ type.

\section{Superfluid Weight}\label{A-D}
\begin{figure}[b]
\centering
\includegraphics[width=0.6\hsize]{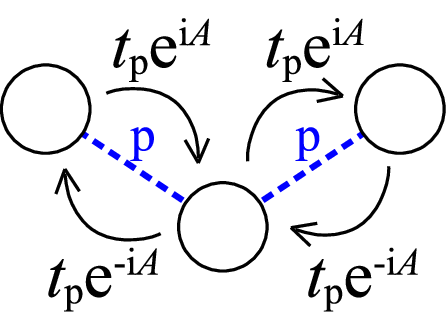}
\caption{\label{Peierls}
Schematic picture of hopping processes with additional phase factors $\text{e}^{\pm\text{i}A}$.
}
\end{figure}

To calculate the superfluid weight, we add the vector potential to the kinetic energy term in the Hamiltonian [Eq.~\eqref{Hamiltonian}] as follows,
\begin{multline}
    H(A)=-\sum_{\left<i,j\right>\sigma}\left[t_{ij}(A)c^{\dg}_{i\sigma}c_{j\sigma}+\text{H.c.}\right] \\
    +U\sum_{i}n_{i\ua}n_{i\da}+\sum_{\left<i,j\right>} V_{ij}n_in_j,
\end{multline}
where $t_p \rightarrow t_p \text{e}^{\pm\text{i}A}$ as shown in Fig.~\ref{Peierls}.
Here, $A$ corresponds to the Peierls phase and induces a current along the $p$ bond.
Using the VMC method, the Drude weight is estimated as
\begin{equation}\label{D}
    D=\left.\frac{\partial^2 E(A)}{\partial A^2}\right|_{A=0},
\end{equation}
where $E(A)$ is a variational energy for $H(A)$.
When the system is insulating, $D=0$ because the energy increase induced by an infinitesimal $A$ is suppressed due to the lack of free carriers, preventing the response to the external field.
However, as pointed out by Millis and Coppersmith~\cite{Millis}, $D$ cannot be exactly zero (i.e., the system appears metallic) as long as the trial wave function is real.

\begin{figure}
\centering
\includegraphics[width=1.0\hsize]{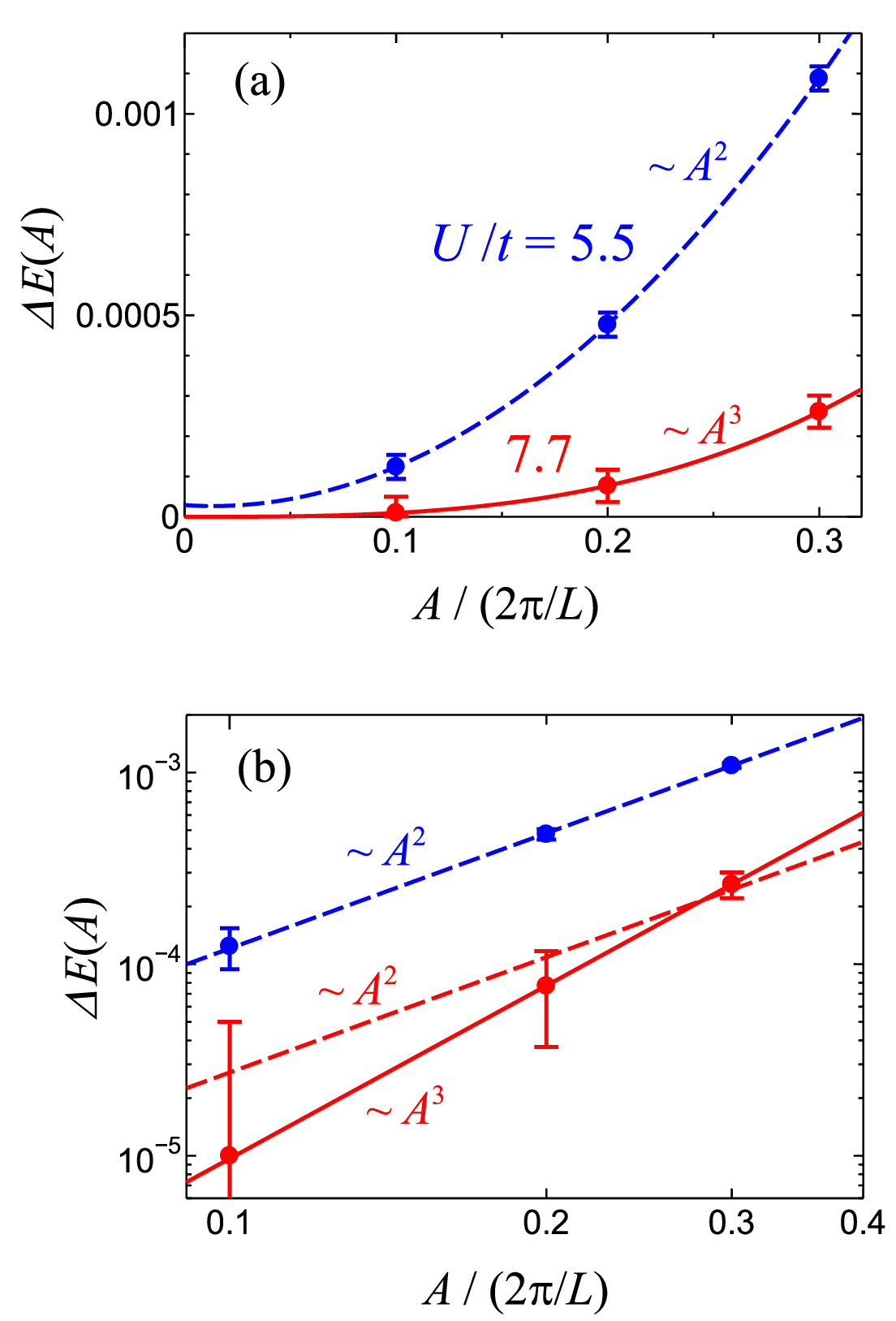}
\caption{\label{EA}
$A$ dependence of the energy increase $\Delta E(A)=E(A)-E(0)$ for $U/t=5.5$ and $U/t=7.7$.
Panel (a) is plotted on a linear scale, while both the vertical and horizontal axes in (b) are plotted on a logarithmic scale.
Solid (dashed) curves represent $A^3$ ($A^2$) fitting.
}
\end{figure}

To address this issue, we introduce a complex phase factor in the trial wave function~\cite{Tamura} defined as
\begin{equation}
    P^{\text{dim}}_{\theta}=\exp\left[ \text{i}\sum_{(m,n)} \theta_{mn}d^{\text{dim}}_m h^{\text{dim}}_n\right],
\end{equation}
where $\sum_{(m,n)}$ runs over nearest-neighbor dimer pair connected by $p$ bonds.
Here, $d^{\text{dim}}_m=n_{m_1\ua}n_{m_1\da}+n_{m_2\ua}n_{m_2\da}$ is the total number of doublons in the $m$-th dimer, and
$h^{\text{dim}}_n=(1-n_{n_1\ua})(1-n_{n_1\da})+(1-n_{n_2\ua})(1-n_{n_2\da})$ is the total number of holons in the $n$-th dimer.
In essence, $P^{\text{dim}}_{\theta}$ introduces a phase to the configuration of neighboring doublon-holon pairs.
We assume a simple form for $\theta_{mn}$ as
\begin{equation}\label{theta}
    \theta_{mn}=
    \begin{cases}
        -\theta & x_m-x_n > 0 \\
        \theta & x_m-x_n < 0
    \end{cases}
\end{equation}
where $\theta$ is a variational parameter.
Through this factor, hopping processes that create or annihilate doublon-holon pairs acquire the complex phase $\text{e}^{\pm\text{i}(A-\theta)}$,
where $\theta$ works to counteract the effect of $A$.

The superfluid density $D_{\text{s}}$ is also estimated using Eq.~\eqref{D} for superconducting trial wave functions~\cite{Scalapino,Tamura,Hetenyi}. 
Figure~\ref{EA} shows the $A$ dependence of the energy increase $\Delta E(A)=E(A)-E(0)$ for $U/t=5.5$ and $U/t=7.7$.
For $U/t=5.5$, $\Delta E(A)$ exhibits a quadratic increase with respect to $A$, and $D_{\text{s}}\approx 0.35$ can be estimated from the coefficient.
In contrast, for $U/t=7.7$, $\Delta E(A)\sim A^3$, and $D_{\text{s}}\rightarrow0$ as $A\rightarrow0$.
Based on the $A$ dependence of $\Delta E(A)$, we conclude that the system is superconducting for $U/t=5.5$, while it is nonsuperconducting and insulating for $U/t=7.7$, despite the fact that the optimized $\tilde{\Delta}_{ij}$ remains finite.
This conclusion is further supported by the superconducting correlation function $P_{\text{SC}}$ in Fig.~2(a).

The reason why the variational phase factor $P^{\text{dim}}_{\theta}$ effectively describes the insulating phase is as follows. 
In the Mott insulating phase, most of the kinetic energy arises from hopping processes that create or annihilate doublon-holon pairs, as these pairs are strongly bound and free carriers are absent. 
Therefore, if $\theta$ is optimized to satisfy $\theta \approx A$, the energy increase induced by $A$ would be nearly canceled out.
Indeed, as shown in Fig.~\ref{EA}, most of the energy increase is canceled for $U/t=7.7$ with the optimized $\theta / A \approx 0.8$.
Although improvements to $P^{\text{dim}}_{\theta}$, such as introducing long-range doublon-holon phase factors, may yield even better results, we believe that the simple form in Eq.~\eqref{theta} captures the essential physics of conducting phenomena in strongly correlated electron systems.

\end{document}